\documentclass{eptcs}

\newcounter{startPage}
\newcommand{\article}[5][]{
  \newcounter{#2}
  \setcounter{#2}{\value{startPage}}
  \addtocounter{startPage}{#3}
}

 \article{fribourgh}{18}
   {Etienne Andr\'e and laurent fribourg}
   {An Inverse Method for Policy-Iteration Based Algorithms}

 \article[(Invited Speaker)]
   {cook}{1}
   {Byron Cook}
   {to be submitted}

 \article[(Invited Speaker)]
   {leucker}{1}
   {Martin Leucker}
   {\emph{Don't know} for Multi-Valued Systems}

 \article{castagna}{5}
   {Giuseppe Castagna, Luca Padovani}
   {A preliminary proposal of decidable testing relations for infinitary asynchronous CCS}

 \article{Tan}{15}
   {Nguyen Tang}
   {A Tighter Bound for the Determinization of Visibly Pushdown Automata}

 \article{bouchy}{20}
   {Florent Bouchy, Alain FINKEL, Pierluigi San Pietro}
   {Dense-choice Counter Machines revisited }

 \article{badban}{8}
   {Bahareh Badban, Jan-Georg Smaus, Stefan Leue}
   {Automated Predicate Abstraction for Real-Time Models}

 \article{morvan}{11}
   {Christophe Morvan}
   {On external presentations of infinite graphs}

\providecommand{\event}{Axel Legay (Ed.): International Workshop on     Verification of Infinite-State Systems (INFINITY)} 

\providecommand{\volume}{13}   
\providecommand{\anno}{2009}   
\providecommand{\firstpage}{1} 
\providecommand{\eid}{1}       

\usepackage[latin1]{inputenc}
\usepackage[english]{babel}

\usepackage{tikz}
\usetikzlibrary{arrows,automata,calc}
\usepackage{amsthm}
\usepackage{amsmath,amssymb,latexsym,enumerate,array,xspace,mathtools}
\usepackage{xcolor}
\usepackage{breakurl}

 \newtheorem{Theo}{\bf Theorem}
 \newtheorem{Prop}[Theo]{\bf Proposition}

 \newtheorem{Coro}[Theo]{\bf Corollary}
\theoremstyle{definition}       

\newtheorem{Defi}[Theo]{Definition}
\newtheorem{Exam}[Theo]{Example}

\newtheorem{Rem}[Theo]{Remark}

\numberwithin{Theo}{section}
\numberwithin{figure}{section} 

\definecolor{aquamarine}{rgb}{0.50,1.00,0.83}%
\definecolor{violetred}{rgb}{0.82,0.13,0.56}%
\definecolor{darkorange}{rgb}{1,0.55,0}%
\definecolor{Orange}{rgb}{1,0.85,0}%
\definecolor{Yellow}{rgb}{1,1,0}%
\definecolor{Peach}{rgb}{1,0.85,0.72}%
\definecolor{Blue}{rgb}{0,0,1}%
\definecolor{Red}{rgb}{1,0,0}%
\definecolor{White}{rgb}{1,1,1}%
\definecolor{Tomate}{rgb}{1,0.1,0.1}%
\definecolor{LemonChiffon}{rgb}{1,.98,.8}%
\definecolor{LemonChiffonDunkel}{rgb}{1,.7,.3}%
\definecolor{VertCite}{rgb}{0.3,0.6,0.5}%
\definecolor{VieuxRouge}{rgb}{0.65,0,0}%
\definecolor{OrangeChiffon}{rgb}{1,0.94,0.88}%
\definecolor{RoseChiffon}{rgb}{1,0.9,0.98}%
\definecolor{Vert}{rgb}{0.3,0.5,0}%
\definecolor{Rose}{rgb}{1,0.3,0.7}%
\definecolor{Framb}{rgb}{0.8,0,0.4}%
\definecolor{BleuC}{rgb}{0.2,0.4,1}%
\definecolor{BleuTC}{rgb}{0.6,0.7,1}%
\definecolor{GrisC}{rgb}{0.7,0.7,0.7}%

\newcommand{\blue}[1]{{\color{blue}#1}}
\newcommand{\green}[1]{{\color{green!60!violet}#1}}

\newcommand{\theocite}[1]{{\bf \cite{#1}}}

\newcommand{\pgd}{\geqslant}
\newcommand{\fifi}{\varphi}
\newcommand{\fiinv}{\varphi ^{-1}}
\newcommand{\nit}{\ensuremath{\mathbb{N}}} 
\newcommand{\vers}{\rightarrow}
\newcommand{\versdans}[1]{\xrightarrow[#1]{}}
\newcommand{\qqs}{\forall}
\newcommand{\inclu}{\subseteq}
\newcommand{\vide}{\emptyset}
\newcommand{\imp}{\Rightarrow}

\newcommand{\chemin}[2]{\xRightarrow[#2]{#1}}               
\newcommand{\prearc}[3]{#1\ v_{#2}\ v_{#3}}                    
\newcommand{\larc}[1]{\xrightarrow{#1}}                    
\newcommand{\arcdans}[2]{\xrightarrow[#2]{#1}}           
\newcommand{\abs}[1]{\lvert#1\rvert}                  
\newcommand{\resp}{{\it resp. }}
\newcommand{\prg}{{prefix-recognizable graph}\xspace}
\newcommand{\prgs}{{prefix-recognizable graphs}\xspace}

\newcommand{\Prgs}{{Prefix-recognizable graphs}\xspace}
\newcommand{\Xbar}{\overline{X}}
\newcommand{\abar}{\overline{a}}
\newcommand{\Abar}{\overline{A}}
\newcommand{\Bbar}{\overline{B}}

\newcommand{\calc}{{\cal C}}

\newcommand{\ens}[1]{\left \{#1\right\}} 
\newcommand{\init}[1]{I}  
\newcommand{\fini}[1]{F}  

\newcommand{\trees}[1]{\mbox{\bf tree}_{#1}} 
\newcommand{\graphs}[1]{\mbox{\bf graph}_{#1}} 
\newcommand{\gegen}{G_{\mbox{gen}}}

\newcommand{\alphab}{\Sigma} 

\newcommand{\sommets}{X} 

\newcommand{\xax}{\sommets^*\times\alphab \times \sommets^*}
\newcommand{\ratg}{Rat(\xax)}

\newcommand{\congg}{{\sc chr}-gram\-mar\xspace}
\newcommand{\tcongg}{tree-{\sc chr}-gram\-mar\xspace}
\newcommand{\tscongg}{tree-separated-{\sc chr}-gram\-mar\xspace}
\newcommand{\hrgg}{{\sc hr}-gram\-mar\xspace}
\newcommand{\hrggs}{{\sc hr}-gram\-mars\xspace}

\newcommand{\labroot}{\text{\bf root}}
\newcommand{\fwd}{\text{\bf fwd}}
\newcommand{\nxt}{\text{\bf nxt}}

\newcommand{\chkA}{\text{\bf chkA}}
\newcommand{\chkB}{\text{\bf chkB}}

\title{On external presentations of infinite graphs}

\author{Christophe Morvan
\institute{Université Paris-Est}
\institute{INRIA Centre Rennes - Bretagne Atlantique\\ 
  Campus de Beaulieu, 35042 Rennes, France}
\email{christophe.morvan@irisa.fr} 
}
\def\titlerunning{On external presentations of infinite graphs}
\def\authorrunning{Christophe Morvan}

\begin{document}

\maketitle



\begin{abstract}
  The vertices of a finite state system are usually a subset of the
  natural numbers. Most algorithms relative to these systems only use
  this fact to select vertices.

  For infinite state systems, however, the situation is different: in
  particular, for such systems having a finite description, each state
  of the system is a configuration of some machine. Then most
  algorithmic approaches rely on the structure of these
  configurations. Such characterisations are said {\em internal}. In
  order to apply algorithms detecting a structural property (like
  identifying connected components) one may have first to transform
  the system in order to fit the description needed for the algorithm.
  The problem of internal characterisation is that it hides
  structural properties, and each solution becomes {\em ad hoc} relatively
  to the form of the configurations.

  On the contrary, {\em external} characterisations avoid explicit
  naming of the vertices. Such characterisation are mostly defined via
  graph transformations.

  In this paper we present two kind of external characterisations:
  deterministic graph rewriting, which in turn characterise regular
  graphs, deterministic context-free languages, and rational graphs.
  Inverse substitution from a generator (like the complete binary
  tree) provides characterisation for prefix-recognizable graphs, the
  Caucal Hierarchy and rational graphs. We illustrate how these
  characterisation provide an efficient tool for the representation of
  infinite state systems.
\end{abstract}

\section{Introduction}

Infinite graphs are a very general way to define infinite state
systems. There are several means to define such infinite graphs:
internal characterisations which relies on some machine: pushdown
systems \cite{muller,Walukiewicz00}, higher order pushdown systems
\cite{cara_wohr}, Petri nets \cite{Petribase}, automatic and rational
graphs \cite{BlGr00,morvan}.  These internal characterisations are
very efficient to prove properties of these systems, but they provide
many restrictions on the names and definition of the states of these
systems. For example, the set of vertices of a rational graph is a
rational set of words, still, having a context-free set of
vertices does not affect the structure of a graph. 

In order to have a more direct access to the structure of such graph
families, external characterisations have been introduced. These
characterisation avoid explicit definition of the vertices.  From a
general perspective these characterisation are based on graph
transformations. Meaning that it is simpler to introduce a suitable
naming for the vertices depending on the problem. Also these
approaches often allow nice proofs for structural properties. 

There are mainly two kind of approach to externally define graph
families: algebraic graph transformation (like inverse rational
substitution) from an original graph (like the complete binary tree).
This technique was first used by Caucal, \cite{caucal}, it allowed him
to prove in a very elegant way the decidability of the monadic second
order of the prefix-recognizable graph (a nice reformulation of this
result in terms of monadic transduction is presented in
\cite{leucker01}). This technique has given rise to the so-called
Caucal hierarchy: \cite{caucal:mfcs}. Graph unfolding (the operation
of transforming a graph into a tree) preserves the decidability of MSO
theory (see \cite{courcelleWalu98}) and Caucal proved that unfolding
\prgs produces trees which are not \prgs, and applying inverse
rational substitution and unfolding alternatively generates a strict
hierarchy of infinite graphs families having decidable MSO theories.
Recently Carayol and Wöhrle showed that this hierarchy coincides in
precise sense to the graphs of higher order pushdown automata:
\cite{cara_wohr}. The rational graphs are a family of infinite graphs
characterising context-sensitive languages, and defined by labelled
rational transducers \cite{morvan, morvan3}.  They are characterised
by similar external characterisation. One of which is by inverse {\em
  finite} substitution from some very general rational graph whose
first order theory is undecidable.

The second kind of external characterisation is done by inductive
graph transformations, and more precisely graph rewriting. The graph
grammars are a classical tool to define infinite families of finite
graphs. In \cite{courcelle}, Courcelle employed deterministic
hyperhedge replacement graph grammars (\hrggs) to define the regular
graphs.  It turns out that these graphs are very close to graphs of
pushdown automata \cite{caucal01}, but enable very elegant proofs for
structural properties like accessibility. \cite{caucal:gragra}
provides very thorough survey for these graphs. Interestingly the
deterministic graphs generated by such grammars correspond precisely
to deterministic context-free languages, this enables a generalisation
of visibly pushdown languages (see \cite{alur04}) defined in
\cite{caucal:sync08}: every deterministic context-free language
belongs to a Boolean algebra of deterministic context-free languages
which contains every regular languages.  Earlier Colcombet,
in \cite{colcomb1}, defined the class of graphs generated by vertex
replacement grammars with product. These graphs have a decidable first
order theory with accessibility. In \cite{morvan:rr09}, the author
introduces contextual graph grammars characterising rational graphs,
and thus context-sensitive languages.

In this paper we propose a detailed survey of these results as well as
a couple of enlightening examples of the interest of working with
external characterisations. The first part of the paper examines graph
families defined from a generator: prefix-recognizable graphs, the
Caucal hierarchy and rational graphs. The second part examines graph
rewriting systems: regular graphs, synchronised graphs and again,
rational graphs.


\section{Preliminaries}

\subsection{Mathematical notations}

For any set $E$, its powerset is denoted by $2^E$; if it is finite,
its size is denoted by $\abs{E}$. Let the set of non-negative integers
be denoted by $\nit$, and $\ens{1,2,3,\ldots ,n}$ be denoted by $[n]$.
A monoid $M$ is a set equipped with an associative operation (denoted
$\cdot$) and a (unique) neutral element (denoted $\varepsilon$). 
A monoid $M$ is {\em free} if there exist a finite subset $A$ of 
$M$ such that $M=A^*:=\bigcup_{n\in\nit} A^n$ and
for each $u\in M$ there exists a unique finite sequence of elements of
$A$, $(u(i))_{i\in [n]}$, such that $u=u(1)u(2)\cdots u(n)$.  Elements
of a free monoid will be called words. Let $u$ be a word in $M$,
$\abs{u}$ denotes the length of $u$ and $u(i)$ denotes its $i$th
letter.

In order to define formally graph grammars, we recall some elements on
hypergraphs. Let $F$ be an alphabet ranked by a mapping $\varrho: F \vers
\nit$, this mapping associates to each element of $F$ its {\em
  arity}. Furthermore, for a ranked alphabet $F$, we denote by $F_n$
the set of symbols of arity $n$. Now given $V$ an arbitrary set, a
{\em hypergraph} $G$ is a subset of $\cup_{n\pgd 1} F_n V^n$.  The
vertex set of such a hypergraph is the set $V_G = \ens{v\in V\ |\
  FV^*vV^* \cap G \neq \vide}$, in our setting, this set is either
finite or countable. A hyperarc of arity $n$ is denoted by 
$f\ v_1\ v_2\ \cdots \ v_n$. 

\subsubsection*{Graphs}
A (simple oriented labelled) {\em graph} $G$ over $V$ with arcs
labelled in $F_2$ is a subset of $F_2VV$.  An element
$ast$ in $G$ is an {\em arc} of {\em source $s$, target $t$ {\em
    and} label $a$} ($s$ and $t$ are {\em vertices} of $G$).  We
denote by $Dom(G)$, $Im(G)$ and $V_G$ the sets respectively of
sources, targets and vertices of $G$. Each arc $ast$ of $G$ is
identified with the labelled transition $ s \arcdans{a}{G} t$ or
simply $ s \larc{a} t$ if $G$ is understood. 

A graph $G$ is {\em deterministic } if distinct arcs with same source
have distinct label: $r\larc{a} s\ \wedge\ r\larc{a} t\ \imp\ s=t $.
The set $2^{{F_2}^+VV}$ of
graphs with vertices in $V$, labelled by elements of ${F_2}^+$, is a
semigroup for the {\em composition relation}:
$G\cdot H:= \{r\larc{a\cdot b} t\ |\ \exists s, r\arcdans{a}{G}
   s\wedge s\arcdans{b}{H} t\}$
   for any $G,H\ \inclu V\times {F_2}^+\times V$.  The relation
   $\arcdans{u}{G^+}$ denoted by $\chemin{\ u\ }{G}$ or simply
   $\chemin{\ u\ }{}$ if $G$ is understood, is the existence of a {\em
     path} in $G$ labelled $u$ in ${F_2}^+$.  A vertex $q\in V_G$ is
   {\em reachable} from a vertex $p\in V_G$ if $p\chemin{\ u\ } G q$
   for some $u\in{F_2}^+$.

   For any subset $L$ of ${F_2}^+$, we denote by
   $s\chemin{L}{} t$ that there exists $u$ in $L$ such that
   $s\chemin{u}{} t$.

   A {\em graph morphism} $g$ is a mapping from a graph $G$ to a graph
   $G'$ such that if there is an arc $u\arcdans a G v$, then there is
   an arc $g(u)\arcdans a {G'} g(v)$. A graph {\em isomorphism} is a
   graph morphism which is a bijection between the vertex sets.

 In the following we will consider first algebraic transformations
 from a generator, then we will examine graphs rewriting system
 defining infinite families of graphs.

\section{Algebraic graph transformations}

\subsection*{Inverse substitution}
A {\em substitution} over a free monoid $X^*$ is a morphism $\fifi :
\alphab^* \vers 2^{X^*}$, which associates to each letter in $\alphab$
a language in $X^*$. For a class $\calc$ of languages in $^*$ (for
example finite languages or regular languages), a $\calc$ substitution
is such that the image of each element of $\alphab$ is a language in
$\calc$.

 We denote by $\Xbar$ the set $\ens{\abar\ |\ a\in
  X}$, and we say that $x\larc{\abar} y$ if $y\larc{a} x$.  Now, given
a graph $G\in XVV$, and $\fifi : \alphab^*\vers 2^{(X\cup \Xbar)^*}$ a
substitution, we define the graph: $\fiinv (G)$ in the following way:
\[
\fiinv ( G )= \{ x\larc{d} y\ |\ d\in\alphab\wedge\  x\chemin{\fifi(d)}{G}y\}
\]

This graph is a subset of $\alphab VV$

\subsection{Prefix-recognizable graphs}

In this section, let $\Lambda$ be the complete binary tree over
$X=\ens{a,b}$, whose vertices are in $V$.

Given a language $L$ in $X^*$, let us denote by $L_{\Lambda}=\{ s\ |\
r\chemin{L}{\Lambda} s \}$, the set of vertices in $\Lambda$ that are
reached by a path in $L$.

\begin{Defi}
  A graph in $\alphab VV$ is prefix-recognizable if it is the image of the complete
  binary tree, $\Lambda$, by an inverse regular substitution followed
  by a regular restriction:
  \[
  \fiinv(\Lambda)_{| L_{\Lambda}}
  \]

  With $\fifi$ a regular substitution ($\fifi:\alphab^* \vers 2^{X^*}$), and $L$ a regular language in $X^*$.
\end{Defi}

\begin{Exam}
  The Figure~\ref{fig:excaucal} represents a classical example of
  \prg, it is an infinite ladder ($L_0$) labelled by $b$'s on the ascending
  side, by $c$'s on the descending side, and with $a$'s connecting the
  ascending and descending branches. 

  On this figure, the complete binary tree $\Lambda$ is composed of $A$
  arcs (dotted) and $B$ arcs (dashed). The graph $L_0$ is obtained
  using a restriction to vertices reached by a path in $L=A+B^*+B^*A$,
  and using the following rational substitution:
  $h(a)=\{A\},\ h(b)=\{B\},\ h(c)=\{\Abar\Bbar A\}$

  For example, in $L_0$, there is an arc labelled $c$, between $BBA$
  and $BA$ (here we identify each vertex of $\Lambda$ with the single
  path from the root leading to it), because there is a path labelled
  $\Abar\Bbar A$ between them. We could also consider the same graph
  with the transitive closure for $c$ arcs, in this case, the
  substitution for $c$ would be: $h'(c)=\{\Abar(\Bbar)^+ A\}$ (which
  is simply the iteration of $h(c)$, simplified by
  $A\Abar=\varepsilon$).

 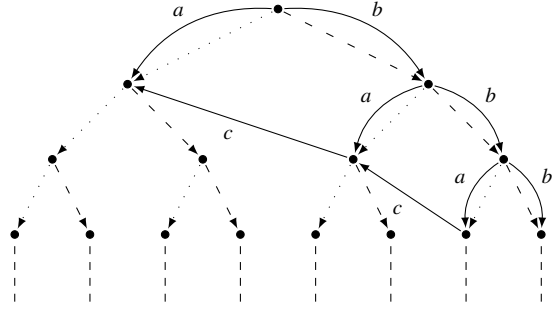
\begin{figure}[h]\centering
         {\scriptsize
         \begin{tikzpicture}[auto, node distance =1cm]
           \tikzstyle{every label}=[label distance= 0.5mm];
           \tikzstyle{sommet}=[draw,shape=circle,inner sep=0pt, outer sep = 0pt, fill= black,minimum size= 1mm];
           \tikzstyle{arcright}=[arrows = -latex, shorten >=1pt, shorten <=1pt, bend right,swap];
           \tikzstyle{arcleft}=[arrows = -latex, shorten >=1pt, shorten <=1pt, bend left];
           \tikzstyle{arct}=[arrows = -latex, shorten >=1pt, shorten <=1pt];
           \tikzstyle{arc0}=[arct, loosely dotted];
           \tikzstyle{arc1}=[arct, loosely dashed];
           \node [sommet] (epsi) {};
            \foreach \fils in {0,1}
            {
              \pgfmathsetmacro{\cote}{(2*\fils)-1}
              \pgfmathsetmacro{\abs}{2*\cote}
              \node [sommet] (\fils) at (\abs,-1){};
              \path[->] (epsi) edge [arc\fils] node {} (\fils);
               \foreach \ffils in {0,1}
               {
                 \pgfmathsetmacro{\cote}{(2*\ffils)-1}
                 \pgfmathsetmacro{\aabs}{\cote}
                 \pgfmathsetmacro{\position}{\abs+\aabs}
                 \def \noeud {\fils\ffils}
                 \node [sommet] (\noeud) at (\position,-2){};
                 \path[->] (\fils) edge [arc\ffils] node {} (\noeud);

                 \foreach \fffils in {0,1}
                 {
                   \pgfmathsetmacro{\cote}{(2*\fffils)-1}
                   \pgfmathsetmacro{\aaabs}{0.5*\cote}
                   \def \nnoeud {\noeud\fffils}
                   \pgfmathsetmacro{\position}{\position+\aaabs}
                   \node [sommet] (\nnoeud) at (\position,-3){};
                   \path[->] (\noeud) edge [arc\fffils] node {} (\nnoeud);
                   \draw [dashed, shorten <=3pt]  (\nnoeud) -- +(0,-1);
                 }
               }      
             }    
             \path[->] (epsi) edge [arcright] node {$a$} (0)
                              edge [arcleft] node {$b$} (1)
                       (1) edge [arcleft] node {$b$} (11)
                           edge [arcright] node {$a$} (10)
                       (11) edge [arcleft] node {$b$} (111)
                           edge [arcright] node {$a$} (110)
                       (10) edge [arct] node {$c$} (0)
                       (110) edge [arct] node {$c$} (10);
         \end{tikzpicture}
     }
 \caption{The infinite ladder $L_0$}
 \label{fig:excaucal}
 \end{figure}
\end{Exam}

\begin{Prop}\theocite{caucal} \label{Prop:caucal-prgs}
  Inverse regular substitution and regular restriction preserves the
  decidability of the MSO theory of graphs.
\end{Prop}

This proposition derives from the inductive definition of regular
languages and MSO formula. From the decidability of the MSO theory of
the complete binary tree we have the following.

\begin{Theo}\theocite{caucal}
  The monadic second order theory of \prgs is decidable.
\end{Theo}

\subsection{Caucal hierarchy}

Another operation preserving monadic second order theory is the
unfolding. Using this operation Caucal has defined a hierarchy of
graphs and terms having decidable MSO theories.

\begin{Theo}\theocite{courcelleWalu98}
  Unfolding preserves the decidability of the MSO theory of graphs.
\end{Theo}

The unfolding of a \prg graph in general produces a graph which is not
a \prg. Let us denote these trees $\trees 2$. Applying inverse regular
substitution followed by regular restriction to these trees produces
graphs (that we denote $\graphs 2$). Iterating this process defines
$\trees n$ and $\graphs n$ for each integer $n\pgd 2$.

All theses graphs (and trees) families are defined by graph
transformations from the complete binary tree. 

By construction the MSO theory of each element in $\trees n$ and
$\graphs n$ is decidable.

And important result is the following:

\begin{Theo}\theocite{caucal:mfcs}
  The hierarchy formed by the $\graphs n$ (\resp $\trees n$) is strict:
  
  \[\graphs n \subsetneq \graphs {n+1} \]
  \[\trees n \subsetneq \trees {n+1} \]
\end{Theo}


This theorem may be summarised in the following table:

\begin{center}
  \begin{tabular}{ccc}
    Level & Trees & graphs \\
    $0$ & Finite trees &  Finite graphs\\
    $1$ & regular trees & \Prgs \\
    $2$ & algebraic trees & $\graphs 2$ \\
    & ... & \\
    $n$ & $\trees n$& $\graphs n$  
  \end{tabular}
\end{center}
This external characterisation corresponds to an internal
characterisation which is higher order pushdown automata: each
$n$-graph is the $\varepsilon$-closure of the {\em configuration
  graph} of a $n$-pushdown automaton (see \cite{cara_wohr}).

\subsection{Context-sensitive languages and rational graphs}

In this section we present an external characterisation for
context-sensitive languages.

\subsubsection{Definitions}

In this section we recall the classical definition of
context-sensitive languages. Then we present the definition of the
family of rational graphs. These graphs are very general, and provide
a {\em graph} characterisation of these languages. More details can be
found in~\cite{morvan,morvan3}.

Context-sensitive languages are defined as the level $1$ of the
Chomsky hierarchy ($0$ being recursively enumerable sets). Which means
they are characterised by growing word grammars. Another popular
characterisation of these languages is
by linear bounded Turing machines \cite{kuroda}.

This family of languages is very expressive, for example, the sets of
words of the form $ww$, or $a^nb^n c^n$, with $n$ a natural number are
context-sensitive sets of words. The set of $a^p$ where $p$ is a prime
number is context-sensitive as well. One of the most stunning property
of these languages is that they are closed under complementation.

The family of rational subsets of a monoid $(M,\cdot)$ is the least
family containing the finite subsets of $M$ and closed under
union, concatenation and iteration.

A transducer is a finite automaton labelled by pairs of words over a
finite alphabet $X$, see for example \cite{berstel}. A
transducer accepts a relation in $X^*\! \times X^*$; these relations
are called rational relations as they are rational subsets of the
product monoid $( X^*\! \times X^*,\cdot)$.


Now, let us consider the graphs of $\xax$. Rational graphs, denoted by
$\ratg$, are extensions of rational relations, which are defined by {\em
  labelled rational transducers}.
\begin{Defi}
  A {\em labelled rational transducer} $T = (Q, \init{T}, \fini{T}, E,
  L)$ over $X$ and $\alphab$, is composed of a finite set of states $Q$, a set of
  initial states $\init{T} \inclu Q$, a set of final states $\fini{T}
  \inclu Q$, a finite set of transitions (or edges) $E \inclu Q\times
  X^*\! \times X^*\!\times Q$ and a mapping $L$ from $\fini{T}$ into
  $2^\alphab$.
\end{Defi}

\begin{figure}[h]
   \begin{tabular}{cr}
     \begin{minipage}[c]{.4\linewidth}
       {\scriptsize
       \begin{tikzpicture}[auto,node distance=2cm]
         \tikzstyle{every edge}=[draw, arrows = -latex];
         \tikzstyle{every loop}=[draw, shorten >=0pt, min distance=3mm];
         \tikzstyle{boucleh}=[out=115, in=65,loop]
         \tikzstyle{etat}=[state,inner sep=2pt,minimum size= 5mm]
         \tikzstyle{istate}=[etat, initial text=, initial]
         \tikzstyle{estate}=[etat, accepting right]
         \draw 
           node[istate] (p) {$p$}
           node[etat] (q1) at +(2.5,1) {$q_1$}
           node[etat] (q2) at +(1,-1.5) {$q_2$}
           node[estate, accepting text=$b$,node distance=4cm] (r2)[right of=p]{$r_2$}
           node[estate, accepting text=$a$] (r1)[above of=r2]{$r_1$}
           node[estate, accepting text=$c$,node distance=1.5cm] (r3)[below of=r2]{$r_3$}
           ;

         \path[->] 
              (p) edge [out=45, in=190] node {$0/0$} (q1)
                       edge node {$0/1$} (r2)
                       edge node {$1/\bot$} (r3)
                       edge [out=80, in=155] node {$\varepsilon/0$} (r1)
                       edge [out=-90, in=145, swap] node {$1/\varepsilon$} (q2)
              (q1) edge [out=-20, in=135] node[pos=.3] {$0/1$} (r2)
              (q2) edge [out=-15, in=195, swap] node {$1/1$} (r3)
              (q1) edge [boucleh] node {$0/0$}()
              (r2) edge [boucleh] node {$1/1$}()
              (q2) edge [boucleh] node {$1/1$}()
              (r1) edge [boucleh] node {$0/0$}();
       \end{tikzpicture}
     }
   \end{minipage}
 &
     \begin{minipage}[c]{.55\linewidth}
       \begin{center}
       {\scriptsize
         \begin{tikzpicture}[auto, node distance =1cm]
           \tikzstyle{every label}=[label distance= 0.5mm];
           \tikzstyle{sommet}=[shape=circle,inner sep=2pt];
           \tikzstyle{arct}=[arrows = -latex, shorten >=2pt, shorten <=2pt];
           \draw
             node [sommet] (epsi) {$\varepsilon$}
             node [sommet] (bot)[below of=epsi] {$\bot$}
             node [sommet] (0) [right of=epsi] {$0$}
             node [sommet] (1) [right of=bot] {$1$};
           \path
            \foreach \source / \but / \lab in {epsi/0/a , 1/bot/c}
            {
              (\source) edge node {$\lab$} (\but)
            }
            (0) edge node {$b$} (1);
           ;
           \foreach \vertex in {0,00}
           {
             \draw
               node [sommet] (0\vertex) at +([shift=(\vertex)] 1, .5) {$0\vertex$}
               node [sommet] (\vertex 1) [below of=0\vertex] {$\vertex 1$};
             \path[->] (\vertex) edge node {$a$} (0\vertex)
                       (0\vertex) edge node {$b$} (\vertex 1);
           }
           \foreach \vertex in {1,11}
           {
             \draw
               node [sommet] (1\vertex) at +([shift=(\vertex)] 1, -.5) {$1\vertex$}
               node [sommet] (0\vertex) [above of=1\vertex] {$0\vertex$};
             \path[->] (1\vertex) edge node {$c$} (\vertex)
                       (0\vertex) edge node {$b$} (1\vertex);
           }
           \path[->] (001) edge node {$b$} (011);

           \draw [dashed, shorten >=5pt] (000) -- +(1,.5);
           \draw [dashed, shorten >=5pt] (111) -- +(1,-.5);

         \end{tikzpicture}
     }
     \end{center}
   \end{minipage}
 \end{tabular}  \centering
  \caption{A rational graph and its labelled transducer}
  \label{fig:anbncn}
\end{figure}
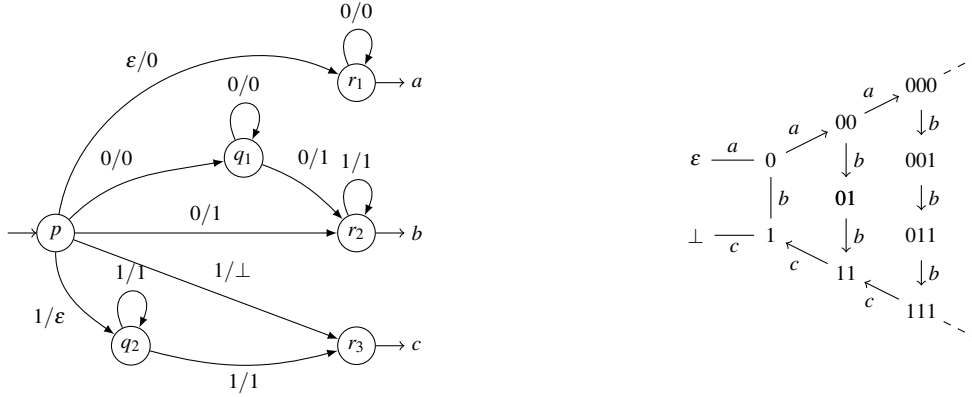

An arc $u\larc{a} v$ is {\em accepted} by a labelled transducer $T$ if
there is a path from a state in $\init{T}$ to a state $f$ in
$\fini{T}$ labelled by $(u,v)$ and such that $a\in L(f)$.
\begin{Defi}\label{carac_grat}
A graph in $\xax$ is {\em rational} if it is accepted by a labelled rational transducer.
\end{Defi}
Let $G$ be a rational graph,
for each $a$ in $\alphab$ we denote by $G_a$ the restriction of $G$
to arcs labelled by $a$ (it defines a rational relation
between vertices); let $u$ be a vertex in $X^*$, we denote by $G_a(u)$ the set 
of all vertices $v$ such that $u\larc{a} v$ is an arc of $G$. 
\begin{Exam}\label{basic}
  In Figure~\ref{fig:anbncn}, the graph on the right-hand side is generated by
  the labelled transducer on the left-hand side.

  The path $p\larc{0/0} q_1 \larc{0/1} r_2 \larc{1/1} r_2$ accepts the
  couple $(001,011)$, the final state $r_2$ is labelled by $b$ thus
  there is a arc $001\larc{b} 011$ in the graph.
\end{Exam}
Rational graphs have been introduced in order to extend existing
families of graphs. They provide a very general family of graphs. They
have few decidable properties, but they characterise context-sensitive
languages \cite{morvan3}. If we only consider trees (rooted connected
acyclic-graphs such that each vertex has at most one predecessor)
these trees have a decidable first order theory \cite{cara-morv}. 

Using transducers to characterise a family of graphs induce that each
graph is defined in a very precise way. In particular, each vertex is
a word, and thus each arc is defined between two precise words, which
are not interchangeable.

Finally, we recall the characterisation of context-sensitive languages
by rational graphs:

\begin{Theo}\theocite{morvan3}\label{theo:csrat}
  The sets of path between regular sets of vertices of rational graphs
  corresponds precisely to context-sensitive languages.
\end{Theo}

\subsubsection{An external characterisation for rational graphs}

There are at least two characterisations of rational graphs in terms
of inverse substitution: the first one is presented in \cite{morvan},
it extends directly Proposition~\ref{Prop:caucal-prgs} to rational
graphs. It uses linear context-free languages, and restricts the use
of symbols of $\overline X$ to left-hand side of productions, and symbols of $X$ to
right-hand side, it is ad-hoc. Here, we present a second such
characterisation build with unrestricted finite
substitutions. Furthermore the generator is no longer the complete
binary tree, but a complex rational graph built on purpose.

\begin{Exam}
  Let $X=\ens{0,1}$ be a fixed alphabet. Let $\gegen$ be the rational
  graph labelled on $X$, defined as follows. First, this graph will be
  used to refine {\em any} graph in $\ratg$, many path in $\gegen$
  will correspond to path in {\em any} transducer. Each state of such
  transducer will be encoded in $X^*$. In fact, $\gegen$ will encode
  $0$ into $000$, $1$ into $001$ for ordinary elements of $X^*$, and
  $0$ into $010$, $1$ into $011$ for elements of $X^*$ which
  represents states. So some vertices will be elements of
  $\ens{000,001}^*$, and some of $\ens{000,001,
    010,011}^*$. Furthermore, $110, 101$ and $100$ will be used to
  mark states, and $111$ will be used along a computation like a
  reading head (and thus some vertices of $\gegen$ will be in $\ens{000,001,
    010,011, 100,101,110,111}^*$).

  Now each pair of elements of $\ens{000,001}^*$ are connected to
  each-other via a infinite set of paths of this form:

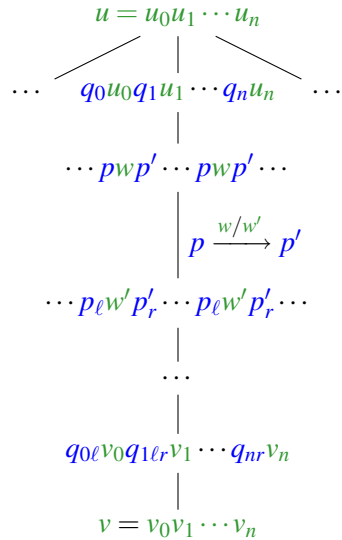
\begin{center}
  \begin{tikzpicture}[auto, node distance =.5cm,level distance=1cm, sibling distance=2cm]
    \node {
      $\green{u=u_0u_1\cdots u_n}$
    }
    child {
      node {
        $\cdots$
      }
    }
    child {
      node {%
        $\blue{q_0}\green{u_0}\blue{q_1}\green{u_1}\cdots \blue{q_n}\green{u_n}$
      }
      child {
        node{%
            $\cdots \blue{p}\green{w}\blue{p'} \cdots \blue{p}\green{w}\blue{p'} \cdots$
        }
        child [level distance=1.8cm]{
          node{%
            $\cdots \blue{p_{\ell}}\green{w'}\blue{p'_{r}} \cdots \blue{p_{\ell}}\green{w'}\blue{p'_{r}} \cdots$
          }
          child [level distance=1cm] { 
            node {
              $\cdots$
            }
            child {
              node{%
                $\blue{q_{0\ell}}\green{v_0}\blue{q_{1\ell r}}\green{v_1}\cdots \blue{q_{nr}}\green{v_n}$
              }
              child {
                node{%
                  $\green v=\green{v_0v_1\cdots v_n}$                    
                }
              }
            }
          }
          edge from parent node[auto=left] {
            $\blue{p}\larc{\green{w}/\green{w'}}\blue{p'}$
          }
        }
      }
    }
    child{
      node {
        $\cdots$
      }
    }; 
  \end{tikzpicture}
\end{center}
The first branching corresponds to {\em non-deterministically}
guessing a path in the transducer that eventually will connect $u$ and
$v$ in the transducer. Each intermediate step corresponds in applying
a transition: each state are copied, except that a special marker is
added: $\ell$ or $r$ representing that the left-hand side
(respectively right-hand side) have been checked (they are encoded by
$110, 101$ and $100$, representing respectively that $\ell$ is
present, $r$ or both). This path is reflected on the labels of the
transitions. Furthermore the marker $111$ is used inside each $pwp'$
to reflect that this part is checked, and also where the progression
is. The last sequence reaching state $v$ is obtained by removing each
state that has been checked.

The key aspect to observe is that along a path of the form
$p\larc{w/w'}p'$ each occurrence of $pwp'$ in the {\em vertex}
$q_0u_0q_1u_1\cdots q_nu_n$ is processed, simultaneously.

Expressed differently: each path in $\gegen$ from a vertex $u$ to a
vertex $v$ reflects the individual transitions of the transducer $T$
such that $(u,v)$ belongs to the rational relation generated by
$T$. And the first intermediate state $q_0u_0q_1u_1\cdots q_nu_n$
reflect an actual {\em path} in $T$, recognising $(u,v)$.
 
\end{Exam}


Now, from this definition of $\gegen$, it is possible to express the
following result:

\begin{Prop}
  Rational graphs are obtained from $\gegen$ by {\em finite}
  inverse substitution and regular restriction.
\end{Prop}
    
\begin{proof}[Proof sketch]
  Given a rational graph $G$ in $\ratg$, and $a$ in $\alphab$, let
  $T_a$ be the rational transducer corresponding to $G_{|a}$. We
  define the finite substitution 

  $h(a)=\ens{p_0\larc{u_0/v_0}p_1\ldots
    p_{m-1}\larc{u_{m-1}/v_{m-1}}p_m\ |\ p_0\in I(T_a)\wedge p_m\in
    F(T_a)\wedge \mbox{{\bf simple}}(p_0\chemin{}{} p_m)}$

  With {\bf simple} denoting the fact that each transition appears at
  most once in the path. $T_a$ is finite, so $h(a)$ is finite. Now,
  from the construction of $\gegen$, each path labelled by an element
  of $h(a)$ connects two vertices which are in the relation defined by
  $T_a$, and thus legitimately connected by an arc labelled $a$.

  To ensure that only vertices not involving "states" are in $G$, we
  add the regular restriction to vertices in $\ens{000,001}^*$.
\end{proof}

Now, from Proposition~2.14 in \cite{morvan} we know that the first
order theory of rational graphs is undecidable. Obviously finite
inverse substitution preserve the decidability of first-order
logic. Thus following Corollary is straightforward.
\begin{Coro}
  First order theory of $\gegen$ is undecidable.
\end{Coro}

Even if the construction of $\gegen$ is built on an explicit naming of
the vertices. This is an external characterisation: every other
rational graph is obtain from an algebraic transformation.

We have seen external characterisation for three families of graph
defined by algebraic transformations from a generator. In the
following section we will examine external characterisations obtained
by recursive graph transformations.

\section{Graph rewriting systems}

In this section we examine graphs defined by graph rewriting systems.
The characterisation are external as they do not provide explicit
naming for the vertices. They are constructed as recursive application
of finite graph transformations.  We focus on \hrgg, and their
contextual counterparts.

\subsection{Deterministic Graph grammars}

Deterministic (hyperhedge replacement) graph grammars are another very
nice example of external characterisation of infinite graphs. These
grammars were initially defined to be an extension to graphs of word
grammars.  Indeed such a graph grammar derived, from an axiom, an
infinite family of {\em finite} graphs. Courcelle in \cite{courcelle}
used the deterministic form of these grammars to obtain a single
infinite graph as the least solution of a finite set of deterministic
graph equations. In 2007 Caucal made a very in-depth survey on
deterministic graphs grammars \cite{caucal:gragra}. In particular he
devised several techniques which allowed the presentation of these
results in a very unified manner.

  \begin{Defi}[Hypergraph grammar]
    A {\em hypergraph grammar} (\hrgg for short) $G$, is a $4$-tuple
    $(N, T, R, H_0)$, where $N$ and $T$ are two ranked alphabets of
    respectively {\em non-terminals} and {\em terminals} symbols;
    $H_0$ is the axiom, a finite graph formed by hyperarcs labelled
    by $N\cup T$, and $R$ is a set of rules of the form $f\ x_1\
    \cdots\ x_{\varrho(f)} \to H$ where $f\ x_1\ \cdots\
    x_{\varrho(f)}$ is an hyperarc joining disjoint vertices and $H$
    is a finite hypergraph.
  \end{Defi}

  \begin{Rem}
    In this paper, we consider graphs, therefore, the terminal symbols
    will have either rank one, or two. Furthermore, we see such a
    graph as a simple subset of $T_2VV\cup T_1V $. Rank $1$ symbols
    will be called {\em colours} rather than labels (we use label to
    identify (hyper) arcs). A single vertex may have several colours.
  \end{Rem}


  A grammar is deterministic if there is a single rewriting rule per
  non-terminal:
  \[
  (X_1,H_1), (X_2, H_2) \in R \wedge X_1(1) = X_2(1) \imp (X_1,H_1) = (X_2, H_2)
  \]

%

  Now, given a set of rules $R$, the {\em rewriting} $\versdans R$ is the
  binary relation between hypergraphs defined as follows: $M$
  rewrites into $N$, written $M\versdans R N$ if there is a
  non-terminal hyperarc $X = Av_1v_2\ldots v_p$ in $M$ and a rule
  $Ax_1x_2\ldots x_p \vers H$ in $R$ such that $N$ is obtained by
  replacing $X$ by $H$ in $M$: $N = (M-X)\cup h(H)$ for some injection
  $h$, mapping $v_i$ to $x_i$ for each $i$, and every other vertices of
  $H$ to vertices outside of $M$. This rewriting is denoted by
  $M\versdans{R,X} N$. Now, this rewriting obviously extends to sets
  of non-terminal, for $E$ such a set, this rewriting is denoted:
  $M\versdans{R,E} N$. The {\em complete parallel rewriting}
  $\chemin{}{R}$ is the rewriting relative to the set of {\em all}
  non-terminal hyperarcs of $R$. 

  Now given a deterministic graph grammar $G=(N, T, R, H_0)$, and a hypergraph $H$,
  we denote by $[H]:= H\cap (T\ V_H\ V_H\cup T\ V_H)$ the set of
  terminal arcs, and colours of $H$. A graph $H$ is {\em generated} by
  $G$, if it belongs to the following set of isomorphic graphs:
  \[
  G^\omega = \ens{\cup_{n\pgd0}[H_n]\ |\ \qqs n\pgd 0, H_n \chemin{}{R} H_{n+1} }
  \]

\begin{Exam}
  We present here a simple example of deterministic graph grammar and
  propose a representation of the resulting graph. An important
  observation on this graph is that it does not provide any naming
  scheme for the vertices. But there is of course an obvious
  connection between the vertices and the sequence of graph rewriting
  producing them. 

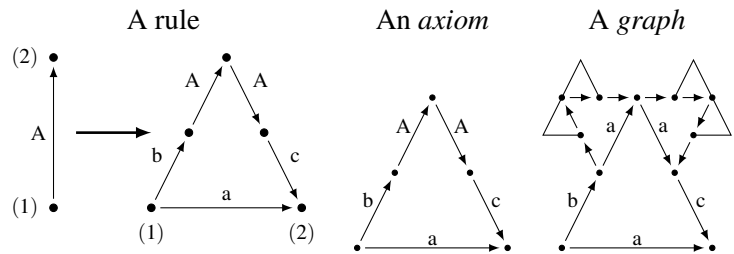
\begin{center}
     \begin{tabular}{ccc}
     A rule & An {\em axiom} & {A {\em graph}}\\
       {\scriptsize
        \begin{tikzpicture}[auto]
          \tikzstyle{every label}=[label distance= 0.5mm];
          \tikzstyle{sommets}=[draw,shape=circle,inner sep=0pt, outer sep = 0pt, fill= black,minimum size= 1mm];
          \tikzstyle{recriture}=[line width = 1pt, arrows = -latex];
          \tikzstyle{arct}=[arrows = -latex, shorten >=2pt, shorten <=2pt];
          \tikzstyle{arcnt}=[arrows = -latex, shorten >=2pt, shorten <=2pt];


          \node (y0) at (0,2) [sommets, label=left:$(2)$]{};
          \node (x0) at (0,0) [sommets, label=left:$(1)$]{}; 
        
          \node (x) at (1.3,0) [sommets, label=below:$(1)$]{};
          \node (y) at (3.3,0) [sommets, label=below:$(2)$]{}; 
        
          \node (b) at (1.8,1) [sommets]{};
          \node (a) at (2.3,2) [sommets]{}; 
          \node (c) at (2.8,1) [sommets]{}; 
        
        
          \draw [recriture] (0.3,1) -- +(1,0);
        
          \draw [arcnt] (x0) -- (y0) node [midway] {A};
          \draw [arcnt] (b) -- (a) node [midway] {A};
          \draw [arcnt] (a) -- (c) node [midway] {A};
        
          \draw [arct] (x) -- (y) node [midway] {a};
          \draw [arct] (c) -- (y) node [midway] {c};
          \draw [arct] (x) -- (b) node [midway] {b};
        \end{tikzpicture}      
        }
     &
       {\scriptsize
        \begin{tikzpicture}[auto]

          \tikzstyle{sommets}=[draw,shape=circle,inner sep=0pt, outer sep = 0pt, fill= black,minimum size= 2pt];
          \tikzstyle{etiqu}=[inner sep=1pt, outer sep = 0pt,minimum size= 2pt];
          \tikzstyle{arct}=[arrows = -latex, shorten >=2pt, shorten <=2pt];
          \tikzstyle{arfin}=[arrows = -, shorten >=0pt, shorten <=0pt];
          \tikzstyle{arcnt}=[arrows = -latex, shorten >=2pt, shorten <=2pt];

        
          \node (x) at (0,0) [sommets]{};
          \node (y) at (2,0) [sommets]{}; 
        
          \node (b) at (.5,1) [sommets]{};
          \node (a) at (1,2) [sommets]{}; 
          \node (c) at (1.5,1) [sommets]{}; 
        
          \draw [arct] (x) -- (y) node [midway,etiqu] {a};
          \draw [arct] (c) -- (y) node [midway,etiqu] {c};
          \draw [arct] (x) -- (b) node [midway,etiqu] {b};

          \draw [arcnt] (b) -- (a) node [midway,etiqu] {A};
          \draw [arcnt] (a) -- (c) node [midway,etiqu] {A};
        \end{tikzpicture}      
        }
     &
       {\scriptsize
        \begin{tikzpicture}[auto]

          \tikzstyle{sommets}=[draw,shape=circle,inner sep=0pt, outer sep = 0pt, fill= black,minimum size= 2pt];
          \tikzstyle{etiqu}=[inner sep=1pt, outer sep = 0pt,minimum size= 2pt];
          \tikzstyle{arct}=[arrows = -latex, shorten >=2pt, shorten <=2pt];
          \tikzstyle{arfin}=[arrows = -, shorten >=0pt, shorten <=0pt];
          \tikzstyle{arcnt}=[arrows = -latex, shorten >=2pt, shorten <=2pt];

        
          \node (x) at (0,0) [sommets]{};
          \node (y) at (2,0) [sommets]{}; 
        
          \node (b) at (.5,1) [sommets]{};
          \node (a) at (1,2) [sommets]{}; 
          \node (c) at (1.5,1) [sommets]{}; 
        
          \draw [arct] (x) -- (y) node [midway,etiqu] {a};
          \draw [arct] (c) -- (y) node [midway,etiqu] {c};
          \draw [arct] (x) -- (b) node [midway,etiqu] {b};

            \draw [arct] (b) -- (a) node [midway,etiqu] {a};
            \node (g1) at (0.25,1.5) [sommets]{};
            \node (g2) at (0.5,2) [sommets]{}; 
            \node (g3) at (0,2) [sommets]{}; 
           
            \draw [arct] (b) -- (g1);
            \draw [arct] (g2) -- (a);

          \draw [arct] (a) -- (c) node [midway,etiqu] {a};

            \node (d1) at (1.75,1.5) [sommets]{};
            \node (d2) at (1.5,2) [sommets]{}; 
            \node (d3) at (2,2) [sommets]{}; 
            \draw [arct] (d1) --  (c);
            \draw [arct] (a) -- (d2);
          \draw [arct] (g3) --  (g2);
          \draw [arct] (g1) -- (g3);
          \draw [arct] (d2) --  (d3);
          \draw [arct] (d3) -- (d1);

          \draw [arfin] (g3) -- ++(.25,.5) -- (g2);
          \draw [arfin] (g3) -- ++(-.25,-.5) -- (g1);

          \draw [arfin] (d3) -- ++(-.25,.5) -- (d2);
          \draw [arfin] (d3) -- ++(.25,-.5) -- (d1);
        \end{tikzpicture}      
        }
   \end{tabular}
\end{center}

\end{Exam}

Graph grammars characterise {\em regular graphs}. This external
characterisation is very efficient to extend to these infinite graphs
techniques which work for finite graphs (for example computing the
connected components of a regular graph is very simple from the
grammar). Furthermore these graphs correspond (in a precise sense) to
transition graphs of pushdown automata. Nonetheless, algorithms
which only depend on the structure of these graphs often make
technical assumptions on the form of the automaton: for example that
the states carry some information, such as the configuration belongs
to a certain regular set. These assumptions only affect the internals
of the automaton, it does not affect the structure of its
configuration graph. In such case, grammars are very efficient as
there is no assumption on vertices identification, only the structure is
explicit.

Following structural operations on graphs preserve the regularity of
graphs, meaning that given a graph grammar $G$ there is an effective
procedure to produce a grammar $G'$ producing the desired graph.

\begin{Prop}
  Accessible colouring preserves regularity.
\end{Prop}

This proposition relies on the fact that there are only finitely many
right-hand side in any grammar so computing local accessibility and
iterating eventually finishes.

\begin{Prop}
  The restriction of a regular graph to vertices having some colour is
  a regular graph.
\end{Prop}

This result is obvious from the definition. And implies that
restriction to a regular set of configuration for a pushdown automaton
is a pushdown automaton, which seems less obvious.

\subsection{Synchronised graph grammars}

These grammars generate deterministic regular graphs. They correspond
to deterministic context-free languages, and enable the extension of
visibly pushdown languages to every deterministic context-free
language. This topic is discussed in \cite{caucal:sync08}. It presents
a nice way to {\em synchronise} deterministic regular graphs.

From this synchronisation, closure properties are defined (mainly
under product) and enables a nice extension of visibly pushdown
automata.

\subsection{Contextual graph rewriting systems}

In this section we present a second external characterisation of
rational graphs. This graph rewriting characterisation is the most
general in some sense: each natural more general rewriting system
fails to produce recursive graphs.

\subsubsection{The general setting}
We recall mainly results from \cite{morvan:rr09}

Let $N_R$ be a finite ranked set of {\em non-terminals}, and $T_R$ a
finite ranked set of {\em terminals}.

We propose here a natural definition of contextual graph rewriting system.

  \begin{Defi}[Contextual graph rewriting system]
    A {\em contextual graph rewriting system} $S$, is a set of rules
    of the form $H_c \cup f\ x_1\ \cdots\ x_{\varrho(f)} \to H_c\cup H$
    where $f\ x_1\ \cdots\ x_{\varrho(f)}$ is a non-terminal hyperarc,
    $H_c$ is a finite {\em context} graph, and $H$ is a finite
    hypergraph, that can share some vertices with $H_c$ and $f$.
    Furthermore, $H_c$ is composed only of terminal hyperarcs, and
    $H_c\cup f\ x_1\ \cdots\ x_{\varrho(f)}$ forms a connected hypergraph.
  \end{Defi}

 \begin{Prop}\label{prop:nonrec}
   Given $(U_i,V_i)_{i\in [n]}$ an instance of PCP, there exists a
   graph obtained from a finite axiom $A$ by a contextual graph
   rewriting system which possesses an arc labelled $\#$ between the
   two vertices $v_0$ and $v_1$ of $A$ if and only if $(U_i,V_i)_{i\in
     [n]}$ is a positive instance.
 \end{Prop}

 The following example illustrates this proposition.

  \begin{Exam}\label{ex-nonrec}
    But the construction is straightforward, and illustrated by this
    example. Consider $((U_i,V_i))_{i\in [n]}$ an instance of PCP, and
    observe the following contextual rewriting system:
    \begin{center}
      {\scriptsize
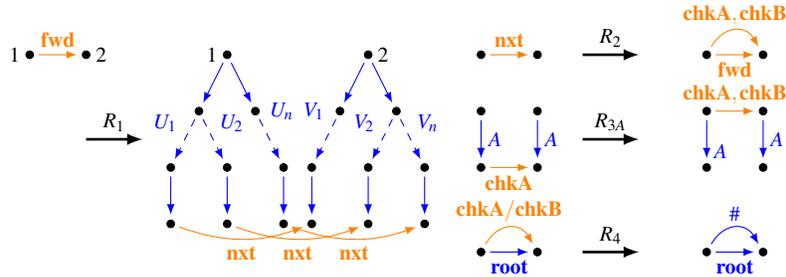
        \begin{tikzpicture}[scale=.75, auto=left]
          %
          
          \tikzstyle{every label}=[label distance= 0.5mm];
          
          \tikzstyle{sommets}=[draw,shape=circle,inner sep=0pt, outer
             sep = 0pt, fill= black, draw=black,minimum size= 1mm];

          \tikzstyle{recriture}=[line width = 1pt, arrows = -latex];

          \tikzstyle{arctx}=[arrows = -latex, shorten >=2pt, shorten
             <=2pt,color=Vert]; 
          \tikzstyle{arct}=[arrows = -latex, shorten >=2pt, shorten <=2pt,color=blue, bend angle=20];
          \tikzstyle{arcnt}=[arrows = -latex, shorten >=2pt, shorten <=2pt,color=orange, bend angle=20];


          \begin{scope}[every node/.style =  sommets]
            \draw (0,3) node (orig1){}
                +(1,0) node (orig2){}
               ++(3.5,0) node (but01){}
                +(2.5,0) node (but02){}
                +(-.5,-1)node (but11){}
                +(.5,-1)node (but12){}
                +(2,-1)node (but13){}
                +(3,-1)node (but14){}
                +(-1,-2)node (but21){}
                +(0,-2)node (but22){}
                +(1,-2)node (but23){}
                +(1.5,-2)node (but24){}
                +(2.5,-2)node (but25){}
                +(3.5,-2)node (but26){}
                +(-1,-3)node (but31){}
                +(0,-3)node (but32){}
                +(1,-3)node (but33){}
                +(1.5,-3)node (but34){}
                +(2.5,-3)node (but35){}
                +(3.5,-3)node (but36){}
                ;
          \end{scope}
          \begin{scope}[node distance=2mm]
            \draw node [left of=orig1] {$1$}
                  node [left of=but01] {$1$}
                  node [right of=orig2] {$2$}
                  node [right of=but02] {$2$}
            ;
          \end{scope}
           \draw[arcnt] (orig1) edge node [midway] {$\fwd$} (orig2)
                        (but31) edge [bend right, auto=right] node [midway] {$\nxt$} (but34)
                        (but32) edge [bend right, auto=right] node [midway] {$\nxt$} (but35)
                        (but33) edge [bend right, auto=right] node [midway] {$\nxt$} (but36)
                        ;
           \draw[arct] (but01) edge node [midway] {} (but11)
                               edge node [midway] {} (but12)
                       (but02) edge node [midway] {} (but13)
                               edge node [midway] {} (but14)
                       (but11) edge [dashed,auto=right] node [midway] {$U_1$} (but21)
                       (but11) edge [dashed] node [midway] {$U_2$} (but22)
                       (but12) edge [dashed] node [near start] {$U_n$} (but23)
                       (but13) edge [dashed,auto=right] node [near start] {$V_1$} (but24)
                       (but14) edge [dashed,auto=right] node [midway] {$V_2$} (but25)
                       (but14) edge [dashed] node [midway] {$V_n$} (but26)
                       (but21) edge node [midway] {} (but31)
                       (but22) edge node [midway] {} (but32)
                       (but23) edge node [midway] {} (but33)
                       (but24) edge node [midway] {} (but34)
                       (but25) edge node [midway] {} (but35)
                       (but26) edge node [midway] {} (but36)
                       ;

          \draw [recriture] (1,1.5) -- +(1,0) node [midway] {$R_1$};
          
          \draw (8,3) node [sommets] (e1){}
               +(1,0) node [sommets] (e2){} 
               +(4,0) node [sommets] (e3){} 
               +(5,0) node [sommets] (e4){} 
              ++(0,-1) node [sommets] (a1){}
               +(1,0) node [sommets] (a2){} 
               +(0,-1) node [sommets] (a3){} 
               +(1,-1) node [sommets] (a4){} 
              ++(4,0) node [sommets] (b1){}
               +(1,0) node [sommets] (b2){} 
               +(0,-1) node [sommets] (b3){} 
               +(1,-1) node [sommets] (b4){} 
              ++(-4,-2.5) node [sommets] (c1){} 
               +(1,0) node [sommets] (c2){} 
               +(4,0) node [sommets] (c3){} 
               +(5,0) node [sommets] (c4){} 
               ;

          \draw [arcnt] (a3) edge node [midway,auto=right] {$\chkA$} (a4)
                        (b1) edge node [midway] {$\chkA,\chkB$} (b2)
                        (c1) edge [out=60, in=120, looseness=1.5] node [midway] {$\chkA/\chkB$} (c2) 
                        (e1) edge node [midway] {$\nxt$} (e2) 
                        (e3) edge [out=60, in=120, looseness=1.5] node [midway] {$\chkA,\chkB$} (e4) 
                        (e3) edge node [midway,auto=right] {$\fwd$} (e4) 
                        ;
          \draw [arct]  (a2) edge node [midway] {$A$} (a4)
                        (a1) edge node [midway] {$A$} (a3)
                        (b2) edge node [midway] {$A$} (b4)
                        (b1) edge node [near end] {$A$} (b3) 
                        (c1) edge node [midway,auto=right] {$\labroot$} (c2) 
                        (c3) edge node [midway,auto=right] {$\labroot$} (c4) 
                        (c3) edge [out=60, in=120, looseness=1.5] node [midway] {$\#$} (c4) 
                        ;

  
          \draw [recriture] (9.8,3) -- +(1,0) node [midway] {$R_2$};
          \draw [recriture] (9.8,1.5) -- +(1,0) node [midway] {$R_{3A}$};
          \draw [recriture] (9.8,-.5) -- +(1,0) node [midway] {$R_4$};

          
        \end{tikzpicture}
      }
    \end{center}
    
    The axiom is simply the following finite graph: $\ens{ \prearc
      \labroot 0 1, \prearc \fwd 0 1}$. Furthermore there is a rule
    $R_{3B}$ similar to $R_{3A}$ for the rewriting of $\chkB$.

    Now, the rule $R_1$ uses arc $\fwd$ to produce two partial binary trees
    corresponding to the $U_i$'s and $V_i$'s. For each sequence of
    indexes $(k_j)_{j\in[m]}$, the extremity of the path
    $(U_{k_j})_{j\in[m]}$ is connected to the extremity of
    $(V_{k_j})_{j\in[m]}$ by an non-terminal arc $\nxt$. Then the
    rules $R_{3A}$ and $R_{3B}$ will ultimately reach the arc
    $\labroot$ if and only if $(U_i,V_i)_{i\in[n]})$ is a positive
    instance of PCP. 
  \end{Exam}

The most direct consequence of this proposition is the following:

\begin{Coro}
  Graphs generated by deterministic contextual graph rewriting systems
  are not recursive. 
\end{Coro}



\subsubsection{Contextual hyper-edge-replacement graph grammars}

In this section we present a more restrictive contextual rewriting
system which will be used to characterise context-sensitive languages.

\begin{Defi}
  A {\em contextual hyper-edge-replacement hypergraph grammar} (\congg for
  short) is a tuple $(C, N, T, R_c, H_0)$, where $C, N$ and $T$ are
  finite ranked alphabets of respectively contextual, non-terminal and
  terminal symbols; $R_c$ is a finite set of contextual rules (for
  each rule $H_c \cup fx_1\dots x_{\varrho(f)} \to H_c\cup H$, the
  graph $H_c$ is formed only by arcs labelled in $C$, and $H$ by arcs
  labelled in $T\cup N$); and $H_0$ is the axiom: a {\em
    deterministic} regular graph formed by arcs with labels in $C$, and
  a single non-terminal hyperarc.
\end{Defi}

This definition imposes that the axiom is a deterministic regular
graph. This restriction ensures that for each rule $R$, of
non-terminal $A$, and each occurrence of $A$ in the graph, there is at
most a single morphism which maps the context of the left-hand side of
$R$ to the neighbourhood of $A$.



First we will show that using a $n$-ary tree as axiom is sufficient
to obtain all the rational graphs up to isomorphism, achieving the
goal of containing the context-sensitive languages.  



\begin{Prop}\label{prop:congg}
  Any rational graph on $\xax$ is obtained from a \congg.
\end{Prop}

\begin{Exam}\label{ex:ratgra}
  Like for Proposition~\ref{prop:nonrec}, the proof is in the full
  paper.  But the construction is straightforward, and illustrated by
  this example. Let $G$ be a rational graph in $\xax$ (and $T$ a
  transducer representing it), let $H_0$ be the complete $n$-ary tree
  labelled on $X$ (with a non-terminal $p_0$ on the root). For each
  state $p$ of $T$, we have the following rule $R_p$.

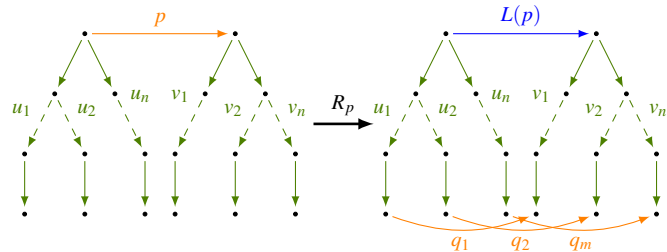
  \begin{center}
\newcommand{\context}[2]{
  \begin{scope}[every node/.style =  sommets]
    \draw (#2)  node (#1a01){}
         +(2.5,0) node (#1a02){}
         +(-.5,-1)node (#1a11){}
         +(.5,-1)node (#1a12){}
         +(2,-1)node (#1a13){}
         +(3,-1)node (#1a14){}
         +(-1,-2)node (#1a21){}
         +(0,-2)node (#1a22){}
         +(1,-2)node (#1a23){}
         +(1.5,-2)node (#1a24){}
         +(2.5,-2)node (#1a25){}
         +(3.5,-2)node (#1a26){}
         +(-1,-3)node (#1a31){}
         +(0,-3)node (#1a32){}
         +(1,-3)node (#1a33){}
         +(1.5,-3)node (#1a34){}
         +(2.5,-3)node (#1a35){}
         +(3.5,-3)node (#1a36){}
         ;
  \end{scope}
  \draw[arctx] (#1a01) edge node [midway] {} (#1a11)
                      edge node [midway] {} (#1a12)
              (#1a02) edge node [midway] {} (#1a13)
                      edge node [midway] {} (#1a14)
              (#1a11) edge [dashed,auto=right] node [midway] {$u_1$} (#1a21)
              (#1a11) edge [dashed] node [midway] {$u_2$} (#1a22)
              (#1a12) edge [dashed] node [near start] {$u_n$} (#1a23)
              (#1a13) edge [dashed,auto=right] node [near start] {$v_1$} (#1a24)
              (#1a14) edge [dashed,auto=right] node [midway] {$v_2$} (#1a25)
              (#1a14) edge [dashed] node [midway] {$v_n$} (#1a26)
              (#1a21) edge node [midway] {} (#1a31)
              (#1a22) edge node [midway] {} (#1a32)
              (#1a23) edge node [midway] {} (#1a33)
              (#1a24) edge node [midway] {} (#1a34)
              (#1a25) edge node [midway] {} (#1a35)
              (#1a26) edge node [midway] {} (#1a36)
              ;
}
      {\scriptsize
        \begin{tikzpicture}[scale=.8, auto=left]
          
          \tikzstyle{every label}=[label distance= 0.5mm];
          \tikzstyle{sommets}=[draw,shape=circle,inner sep=0pt, outer sep = 0pt, fill= black,minimum size= 0.5mm];
          \tikzstyle{recriture}=[line width = 1pt, arrows = -latex];
          \tikzstyle{arct}=[arrows = -latex, shorten >=2pt, shorten <=2pt,color=blue, bend angle=20];
          \tikzstyle{arcnt}=[arrows = -latex, shorten >=2pt, shorten <=2pt,color=orange, bend angle=20];
          \tikzstyle{arctx}=[arrows = -latex, shorten >=2pt, shorten <=2pt,color=Vert, bend angle=20];
          
          \context{left}{0,3}

          \draw [recriture] (3.8,1.5) -- +(1,0) node [midway] {$R_p$};
    
          \context{right}{6,3}
           \draw[arcnt] (lefta01) edge node [midway] {$p$} (lefta02)
                        (righta31) edge [bend right, auto=right] node [midway] {$q_1$} (righta34)
                        (righta32) edge [bend right, auto=right] node [midway] {$q_2$} (righta35)
                        (righta33) edge [bend right, auto=right] node [midway] {$q_m$} (righta36)
                        ;
          \draw [arct]  (righta01) edge node [midway] {$L(p)$} (righta02);
        \end{tikzpicture}
      }
  \end{center}

  Here, we suppose that there are transitions $p\larc{u_i/v_i} q_i$
  for some states $(q_i)_{i\in [m]}$, and also $L(p)$ represent all
  labels produced at state $p$ (if $p$ is a terminal state). Now each
  pair of path in $H_0$ correspond to a pair of paths in $T$.  Thus
  the graph obtained from the contextual rewriting system is the same
  as the graph obtained from the transducer.
\end{Exam}

\begin{Coro}\label{coro:inclu1}
  Any context-sensitive language $L$ is the set of paths between two
  colours in a graph obtained from a \congg.
\end{Coro}

\subsubsection{Graphs obtained from a tree-separated contextual grammar
  are rational graphs}

In this section we examine restrictions in order to obtain a converse
to Proposition~\ref{prop:congg}.

First, we designate interesting restrictions of \congg.  A \congg $(C,
N, T, R_c, H_0)$ is called a {\em \tcongg} if the axiom $H_0$ is a
tree, and left-hand side of each rule of $R_c$ is formed by trees
{\em rooted} in the vertices of the non-terminal (some vertices of this
non-terminal may be non-root vertices of theses trees). Furthermore, if
each such tree possesses a single vertex belonging to the non-terminal
(its root) this grammar is called a {\em \tscongg}.
These grammars are captured by rational graphs:

\begin{Prop}\label{prop:converse1}
  Any graph obtained from a \tscongg, is isomorphic to a rational
  graph on $\xax$.
\end{Prop}

Now combining this result with Theorem~\ref{theo:csrat} and
Corollary~\ref{coro:inclu1} we obtain the desired result.
\begin{Theo}
  The set of paths (between colours) of any graph obtained from a
  \tscongg, is a context-sensitive language. And conversely, any
  context-sensitive language can be obtained as the set of paths of
  such a graph.
\end{Theo}

Now we show that the natural extension of the previous result by
allowing the non-terminal (of the left-hand side) to be set anywhere
in the context produces another non-recursive family of graphs.

\begin{Prop}\label{prop:confails}
  There is a graph obtained from a \congg, such that the axiom is a
  deterministic tree, and having a loop on the root of the axiom if
  and only if a given instance of PCP has a solution.
\end{Prop}

Unfortunately, at the moment, there are few applications illustrating
the potential of this characterisation. A nice one, would be to
provide a new demonstration of the closure under complementation of
context-sensitive languages. Unfortunately the most obvious proof of
this result would require determinism for these graphs, and we have
some indications that deterministic rational graphs do not
characterise all context-sensitive languages.

\section{Discussion}

In this paper we have presented several external characterisations of
infinite graphs families. These characterisations falls into two
categories: either algebraic transformations from a generator, or
recursive application of finite graph transformations.

Our statement is that these characterisations are essential in order
to grasp structural properties of graphs. And also provide an elegant
way to extend to infinite graphs techniques used for finite graphs.
In particular \hrgg enable many simplifications in proofs relatively
to those using pushdown automata.



\end{document}